\documentclass[preprint]{aastex}

\def\lessim{\mathrel{\hbox{\rlap{\hbox{\lower4pt\hbox{$\sim$}}}\hbox{$<$}}}}
\def\grtsim{\mathrel{\hbox{\rlap{\hbox{\lower4pt\hbox{$\sim$}}}\hbox{$>$}}}}

\shorttitle{NOVAE IN NGC 2403}
\shortauthors{Franck et al.}

\begin{document}
\title{The Nova Rate in NGC 2403}


\author{J. R. Franck\altaffilmark{1}, A. W. Shafter\altaffilmark{1}, K. Hornoch\altaffilmark{2}, K. A. Misselt\altaffilmark{3}}
\altaffiltext{1}{Department of Astronomy, San Diego State University, San Diego, CA 92182, USA}
\altaffiltext{2}{Astronomical Institute, Academy of Sciences, CZ-251~65~Ond\v{r}ejov, Czech Republic}
\altaffiltext{3}{Steward Observatory, University of Arizona, Tucson, AZ, 85721, USA}

\begin{abstract}
A multi-epoch H$\alpha$ survey of the late-type spiral galaxy NGC 2403
has been completed in order to determine its nova rate. A
total of nine nova candidates were discovered in 48 nights of observation
with two different telescopes over the period from 2001 February
to 2012 April. After making corrections for temporal
coverage and spatial completeness, a nova rate of
2.0$^{+0.5}_{-0.3}$ yr$^{-1}$ was determined. This rate
corresponds to a luminosity-specific nova
rate of $2.5\pm0.7$ novae per year per 10$^{10} L_{\odot,K}$.
This value is consistent with that of the similar
Hubble type galaxy, M33, and is typical of those of other galaxies
with measured nova rates, which range from $1-3$ novae per
year per 10$^{10} L_{\odot,K}$.
\end{abstract}

\keywords{galaxies: individual (NGC 2403) --- novae, cataclysmic variables}

\section{Introduction}

Population synthesis models have predicted the infrared (e.g., $K$ band) 
luminosity-specific nova rate (LSNR)
of a galaxy should be sensitive to its star
formation history, with younger stellar populations producing nova
eruptions at a higher rate \citep{yun97}. This dependence is the result of
an inverse correlation between the average mass of the white dwarfs
in nova progenitor binaries and the
time elapsed since the zero-age main sequence
system formed \citep{tut95}. Since novae harboring
higher mass white dwarfs are expected to exhibit
more frequent, and more luminous, eruptions
\citep{rit91, liv92, kol95}, galaxies that form most of their stars
in an early burst of star formation (i.e., ellipticals), were not
expected to be prolific nova producers. On the other hand, galaxies that
have experienced a more uniform star formation rate over time, such as
late-type spiral galaxies, were predicted to have higher LSNRs.

In recent years, nova surveys have been conducted
in galaxies spanning a range of Hubble types, providing the
opportunity to study novae in isolated stellar populations \citep{sha08}.
\citet{cia90a}
first published a study of galaxies with measured nova rates from which they
concluded that the $K$-band LSNR was basically
independent of Hubble type.
Several years later, however,
\citet{del94} re-analyzed many of these same galaxies and
concluded that the LSNRs are
systematically higher in the late-type,
low-mass galaxies compared with their earlier type
counterparts.
Since these first two pioneering studies, several additional nova
surveys have been conducted \citep{sha02, fer03, wil04, coe08, gue10}.
Like \citet{sha00}, most of these studies also failed to find
compelling evidence that the LSNR varies significantly with Hubble type.

To make further progress in determining how sensitive the LSNR may be to
the properties of the underlying stellar population, it is important to
obtain nova rates for additional galaxies spanning a broad range of
Hubble types. Here we report the results of a multi-year survey
to determine the nova rate in the late-type spiral galaxy NGC 2403.

\section{Observations}

NGC 2403 was observed over a total of 48 nights from 2001 February
until 2012 April, as summarized in Table~1.
The data were collected with two telescopes:
the Steward Observatory's
2.3-m Bok Telescope (BT) and the 2.5-m Isaac Newton Telescope (INT).
The BT observations were obtained with the 90Prime
Camera \citep{will04},
which contains a mosaic of four 4K $\times$ 4K CCDs at prime focus.
The INT observations were made with the Wide Field Camera (WFC),
which contains a mosaic of four thinned EEV 2K x 4K CCDs at prime focus.
The CCDs
have a pixel size of 13.5 microns, which
corresponds to a plate scale of $\sim$0.33 arcsec/pixel. The edge
to edge limit of the mosaic, neglecting the ~1 arcmin inter-chip spacing,
is 34 arcmins. We used archival images stored in the Isaac Newton Group Archive
as well as recent images taken during three D-nights in 2011--2012.

As in previous nova surveys, we chose to image NGC 2403
in H$\alpha$. Imaging in H$\alpha$ offers several advantages over
broad-band observations \citep{cia90b}.
Novae develop strong H$\alpha$ emission lines shortly
after eruption that persist long after the continuum has faded. As a result
novae are more easily detected in synoptic surveys like ours.
Secondly, the
H$\alpha$ images provide a greater contrast against the bright
galaxy background than do broad-band observations. The greater contrast
results in increased survey completeness in the bright central regions of
galaxies. Finally, H$\alpha$ observations are less affected by
extinction internal to the target galaxy than are broad-band observations,
which are typically taken at shorter wavelengths.
The BT observations employed a narrow-band H$\alpha$ filter centered at
6580 \AA\ with a FWHM of $\sim$80 \AA , while the INT data were taken
through a similar H$\alpha$ filter having a 95 \AA\ FWHM
centered at 6568 \AA. The slight differences in the characteristics
of the H$\alpha$ filters do not significantly affect our results
given the intrinsic variation in nova H$\alpha$ emission line widths.

For the INT images, standard reduction procedures for raw CCD
images were applied (bias subtract and
flat-field correction) using APHOT,
a synthetic aperture photometry and astrometry
software developed by M. Velen and P. Pravec
at the Ond\v{r}ejov observatory
\citep[see][]{pra94}, and SIPS\footnote{\tt http://ccd.mii.cz/} programs.
Reduced images of the same epoch were co-added to improve the signal-to-noise
ratio.
The gradient of the galaxy background
of co-added images was flattened by the spatial
median filter using SIPS.

Data obtained with the BT were first processed using custom software. 
Since NGC 2403 fell on a single CCD of the four CCD 90prime camera, only 
the CCD containing NGC 2403 was processed.  Given that each CCD is read out by
two amplifiers, with the data stored separately, the individual 
amplifier images were first orientated properly and stitched together to 
create the final image. Known bad pixels/columns were masked prior to further
analysis. After re-constructing each image, roughly 10--20
bias frames were combined into a single average bias.  Similarly, 10--20
dark frames with an exposure time matched to the object exposure time were
combined into an average dark.  Five to ten twilight sky flats were averaged
to construct the appropriate flat.  Object data were bias subtracted, dark
subtracted, and flat fielded using the average calibration frames. A rough 
coordinate system was assigned to each image given the known pointing and 
knowledge of the focal plane geometry.   

Following these basic reduction steps, a WCS system was assigned to each 
individual frame using subpackages of the IRAF task MSCRED. Using the rough
coordinate system defined above, a catalog of bright sources was generated
and overlaid on the image. Using MSCTPEAK, 30--40 of the catalog sources were
associated with counterparts on the image.
A fourth- or fifth-order polynomial was then fit 
to minimize the residuals and used to transform
all catalog sources to the image system.
After the distortion coefficients were determined as described,
each individual image was projected
onto a regular grid centered on the nominal coordinates of NGC 2403 using
MSCIMAGE. For a given epoch, all projected images were combined using
MSCSTACK, and for comparison between epochs, WREGISTER was used to align each 
epoch to a common system. 

\subsection{Nova Detection}

Nova candidates were selected based on their observed brightness and
the nature of their variability. Novae are transient variables
with a rapid rise to peak luminosity ($-10\lessim M_V\lessim -6$) followed
by a slower return to quiescence with a timescale that varies significantly
from nova to nova. It is often challenging to differentiate low-luminosity
novae from long-period variables (LPVs), such as Mira variables, which
have similar luminosities. Unlike novae, however, LPVs are
sources with a limited magnitude range
that can be detected multiple times in a survey
having a sufficiently deep limiting magnitude
and sufficiently long time coverage. Thus,
to qualify as a nova, a variable source had to appear
suddenly (i.e., not be detected in prior epochs) and then fade
(or disappear completely) in subsequent epochs. The fields of
all nova candidates were scrutinized
to make sure that the nova candidates were
undetectable in all epochs observed more than a year from the date
of discovery.
For eliminating the possibility that some of our faint
nova candidates could be in fact red LPVs,
we also used $V$, $R$, and $I$-band images of NGC 2403 taken with the
Suprime-Cam mounted on the 8.2-m Subaru
telescope and obtained from the SMOKA \citep{bab02}.
Using these images we rejected a few faint nova candidates,
due to their presence on images obtained well
before or after the H$\alpha$ detection and their apparently red color, which
suggests that they are red LPVs around maximum light.

For the images obtained with the BT,
identification of the candidates in the crowded
and complex background of NGC 2403 was expedited
using the ISIS package \citep{ala98},
which aligns the images over all epochs and differences them.
The resulting images are subsequently blinked
by eye to reveal variable sources. Of the several transient sources
discovered, two passed our criteria for inclusion as novae:
N2403~2005-10a and 2005-12a.
NGC 2403 is host to a number of H \texttt{II} regions,
which are bright in H$\alpha$ and are
difficult to cleanly difference with ISIS.
In the attempt to recover possible candidates in these regions,
as well as the nucleus of the galaxy,
the IRAF \texttt{median} routine was performed
using a variety of box sizes,
with a 9 $\times$ 9 pixel box producing the cleanest images.
With the bright background light of NGC 2403,
blinking the median-subtracted images
further enhanced the completeness of our survey.
This process, however, did not produce any additional nova candidates.

The nova candidates from
the INT images were identified through a visual comparison of
a given image with a high-quality master image after the images had
been flattened by median filtering.
Master images, which
were created from images separated by at least three years from
the epoch being searched, were obtained under
very good seeing conditions that resulted in deeper
limiting magnitudes compared with the searched image.
For comparison with the deepest images, we used
a master image created from the combination of more than one epoch.
Once transient sources were identified, their transient nature
was confirmed through a comparison with
all other INT images and the Subaru images.
Finally, the confirmed transients found in the INT images
were cross-checked against our BT images to assure that they
were not visible at other epochs. A similar cross-check
was performed for the BT candidates using the INT images.
Although our nova search method might seem overly simplistic,
we have found that it does a better job in detecting transient sources
in a galaxy such as NGC 2403 where the complex structures of H$\alpha$ regions
renders more sophisticated and automated detection methods less effective
in the discovery of transient point sources.

\subsection{Nova Astrometry and Photometry}

Astrometry of novae discovered in the INT data
was conducted in APHOT using positions of field stars from the UCAC2
\citep{zac04}.
Derived positions of novae have uncertainties in order of tenths
of an arcsecond.
Astrometry of the two BT novae, as well as
photometry of all novae in our survey, was carried out with the aid of the
IRAF \texttt{phot} routine in the APPHOT package.
Magnitudes for the novae were then determined through a differential
comparison with respect to three secondary H$\alpha$ standard stars.
The secondary standard stars,
N2403-S1 ($\alpha_\mathrm{J2000} = 07^h36^m04\fs67, \delta_\mathrm{J2000} = +65\degr34\arcmin39\farcs66$),
N2403-S2 ($\alpha_\mathrm{J2000} = 07^h36^m11\fs78, \delta_\mathrm{J2000} = +65\degr35\arcmin55\farcs41$),
N2403-S3 ($\alpha_\mathrm{J2000} = 07^h37^m31\fs98, \delta_\mathrm{J2000} = +65\degr38\arcmin47\farcs09$),
were calibrated with respect to the photometric standard Feige 34 \citep{sto77},
and found to have $m_{H\alpha}$ = 15.18, $m_{H\alpha}$ = 16.00,
and $m_{H\alpha}$ = 16.46, respectively.
As in our previous surveys \citep{sha01,wil04,coe08,gue10},
we have assumed an 100\% filling factor for the H$\alpha$ filter.
This assumption allows consistency between surveys,
with the recognition that the bandpass will not always be filled
by a particular nova's H$\alpha$ emission.
For computing of offsets of particular novae from the NGC 2403 center we used
$\alpha_\mathrm{J2000} = 7^h36^m51\fs40,
\delta_\mathrm{J2000} = +65\degr36\arcmin09\farcs2$
as the reference position for the center of NGC 2403.

Table 2 contains the dates, positions,
and $m_{H\alpha}$ of all nine candidates found in our survey.
Each candidate was detected in at least one subsequent epoch.
Two of the candidates, N2403 2001-02a and N2403 2010-02a,
were found to have very slow fade rates of greater than 6 months,
indicating that these are particularly slow novae.

\section{The Nova Rate in NGC 2403}

\subsection{The Survey Completeness}

An estimate of the nova rate in NGC 2403 requires a knowledge of
the completeness of the survey as a function of magnitude.
Unlike our previous surveys
\citep[e.g.,][]{coe08,gue10}, the present study has been conducted using
different telescopes, instruments, and observing conditions. As a result,
the limiting magnitude of our observations varies significantly
over each epoch of observation. To account for this variability
the completeness of each epoch should be assessed separately. In our
earlier studies, we have determined the survey completeness
by performing artificial star tests using the IRAF routine \texttt{addstar}.
Rather than conduct artificial star tests on each of the 48
nights individually (which would be impractical), we have chosen to
perform the artificial star tests on a single ``fiducial" image
and then shift this completeness function by an amount $\Delta m_i$
appropriate for images obtained in
each of the remaining epochs. The value of $\Delta m_i$
($= m_{lim,0} - m_{lim,i}$) is simply the difference in
magnitude between the faintest star that can be reliably detected
in our fiducial epoch and that in the $i$th epoch.

\begin{figure}
\includegraphics[angle=-90,scale=0.65]{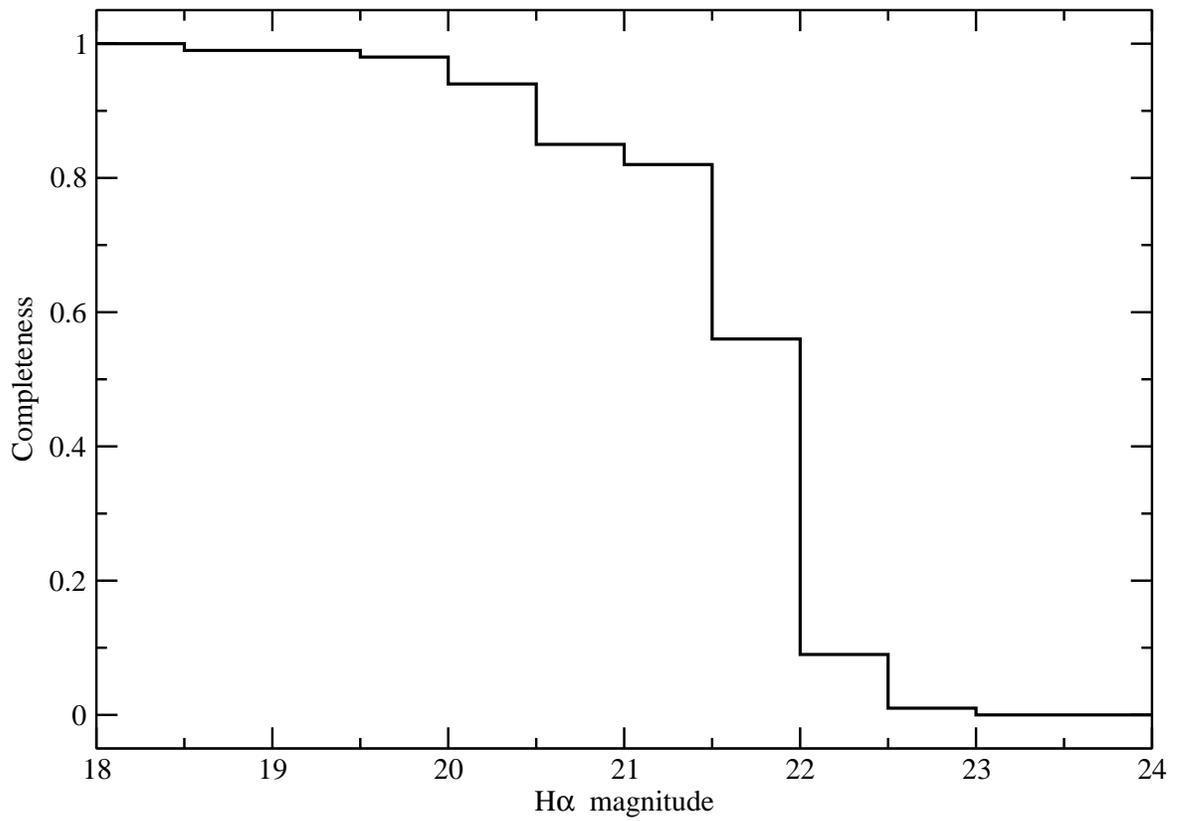}
\caption{The completeness function, $C(m)$, for our fiducial image showing
 the fraction of test novae recovered in our artificial star tests
 as a function of H$\alpha$ magnitude.}
\label{fig:LimitingMag}
\end{figure}

We began the completeness tests by creating
artificial novae with magnitudes ranging between $m_{H\alpha}$ = 18.0 and
$m_{H\alpha}$ = 24.0, which were then divided into 12
equally-spaced 0.5 mag bins. Next, following the spatial distribution of the
$K$-band light [from the data of \citet{pie97}],
100 artificial novae in each magnitude bin were
randomly distributed throughout the fiducial image, which we take
to be the BT image from 2005 December 28.
We searched for these artificial ``novae'' following the same
procedures as the ones employed to discover the actual novae. In practice,
however, our detection criteria were somewhat different.
In conducting our artificial
star tests, the completeness at a given magnitude was simply taken
to be the fraction of artificial novae recovered.
On the other hand, for a transient source to qualify
as an actual nova candidate, we required that it
be $\sim0.5$ mag brighter than that required for a mere detection.
To compensate for this difference in detection criteria,
we have applied a corresponding 0.5 mag correction to each magnitude bin
of the completeness function.
This correction renders the effective limiting magnitude
of the fiducial epoch 0.5 mag brighter than it would be if based directly
on the artificial star tests.
The resulting completeness function for the fiducial image, $C(m)$,
is shown in Figure \ref{fig:LimitingMag}. For this
epoch, the recovered nova fraction starts to decline steeply at magnitude of
$m_{H\alpha}\simeq21.5$. The completeness function for the $i$th epochs can
then be simply estimated as $C(m+\Delta m_i)$.

\subsection{The Monte Carlo Procedure}

\begin{figure}
\includegraphics[angle=-90,scale=0.65]{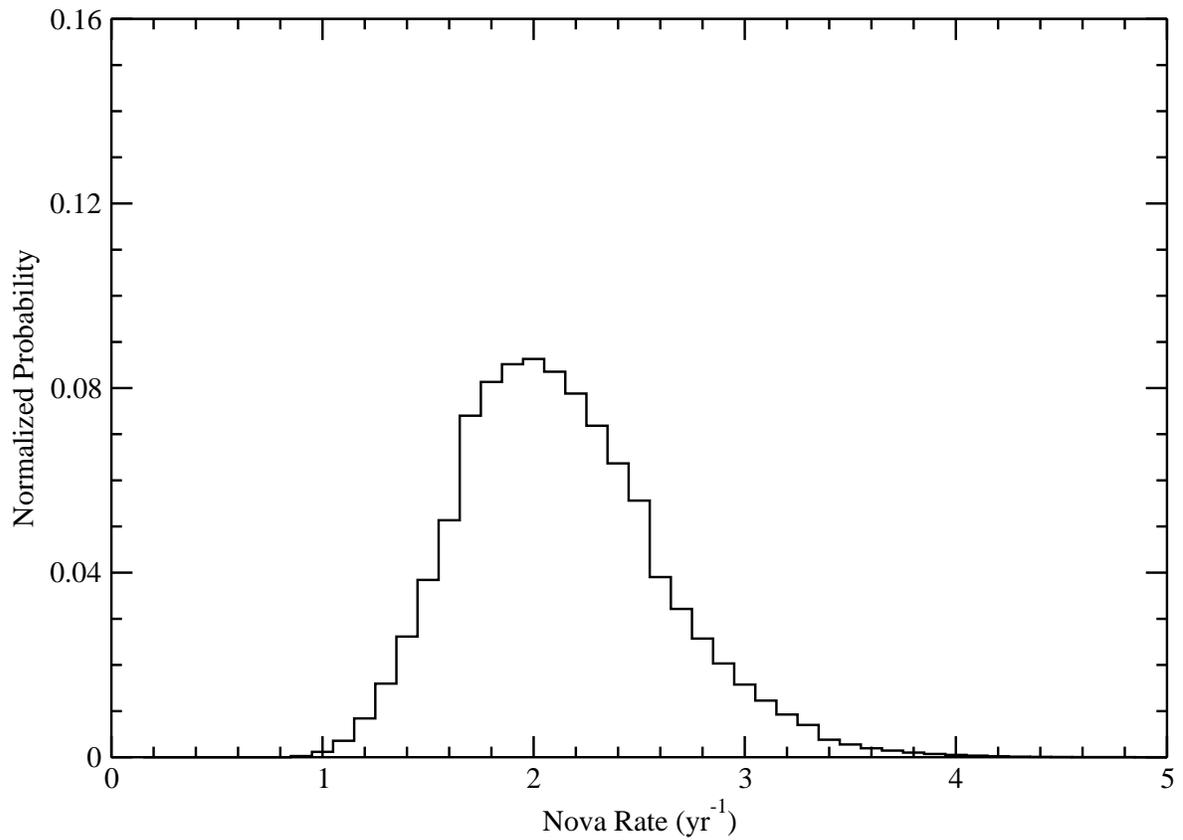}
\caption{Results from the Monte Carlo simulation for the nine nova candidates.
 The peak in the normalized probability distribution
 ($R=2.0^{+0.5}_{-0.3}$~yr$^{-1}$) represents the most probable
  nova rate in NGC 2403.}
\label{fig:MonteCarlo}
\end{figure}

We have estimated the nova rate in NGC 2403
by employing a Monte Carlo simulation similar to that we applied
previously for M101 and M94 \citep{coe08,gue10}. As just described,
a difference
in our approach here is that we have explicitly taken into account
differences in completeness at various epochs.
In the current analysis, we compare
the number of novae discovered in our
survey ($n_{obs}$ = 9)
with an estimate of the number of novae we would expect to see,
$N_{obs}$, which is a function of our survey depth, temporal sampling,
and of course, $R$, the intrinsic nova rate in NGC 2403.
We begin by computing, for a wide range of possible $R$,
a set of model H$\alpha$ light curves based on randomly
selected peak magnitudes and decay rates for
M31 and M81 novae \citep{sha01,nei04}. Then, based on the
dates of our survey (see Table~1) and our adopted distance to NGC 2403
[$\mu_{0} = 27.48 \pm 0.10$ \citep{fre01}],
we computed an observed nova luminosity
function for each epoch, $n_i(m,R)$,
over the complete range of possible intrinsic nova rates, $R$.
Finally, the resulting
luminosity function was convolved with the completeness function,
$C(m+\Delta m_i)$, appropriate for the $i$th epoch, and then summed
over all epochs to determine the expected number of novae in NGC 2403: 

\begin{equation}
N_{obs}(R) = \sum_{i}\sum_{m}{C(m+\Delta m_i)~n_i(m,R)},
\label{eqn:NovaNum}
\end{equation}

\noindent
where $\Delta m_i = m_{lim,0} - m_{lim,i}$.
As part of a Monte Carlo process,
we repeated the above procedure 10$^{5}$ times over the range
$1<R~(\mathrm{yr}^{-1})<10$
in steps of $0.1~\mathrm{yr}^{-1}$, and recorded the number
of times that $N_{obs}(R)$ matched $n_{obs}$ = 9, the actual number of novae
identified in this survey. After normalizing the number of matches as a
function of $R$, we obtained the probability distribution function shown in
Figure \ref{fig:MonteCarlo}. The most probable nova rate is represented by the
peak in the distribution at 2.0$^{+0.5}_{-0.3}$ yr$^{-1}$, where the error
range encompasses 50\% of the integrated probability distribution.

\begin{figure}
\includegraphics[angle=-90,scale=0.65]{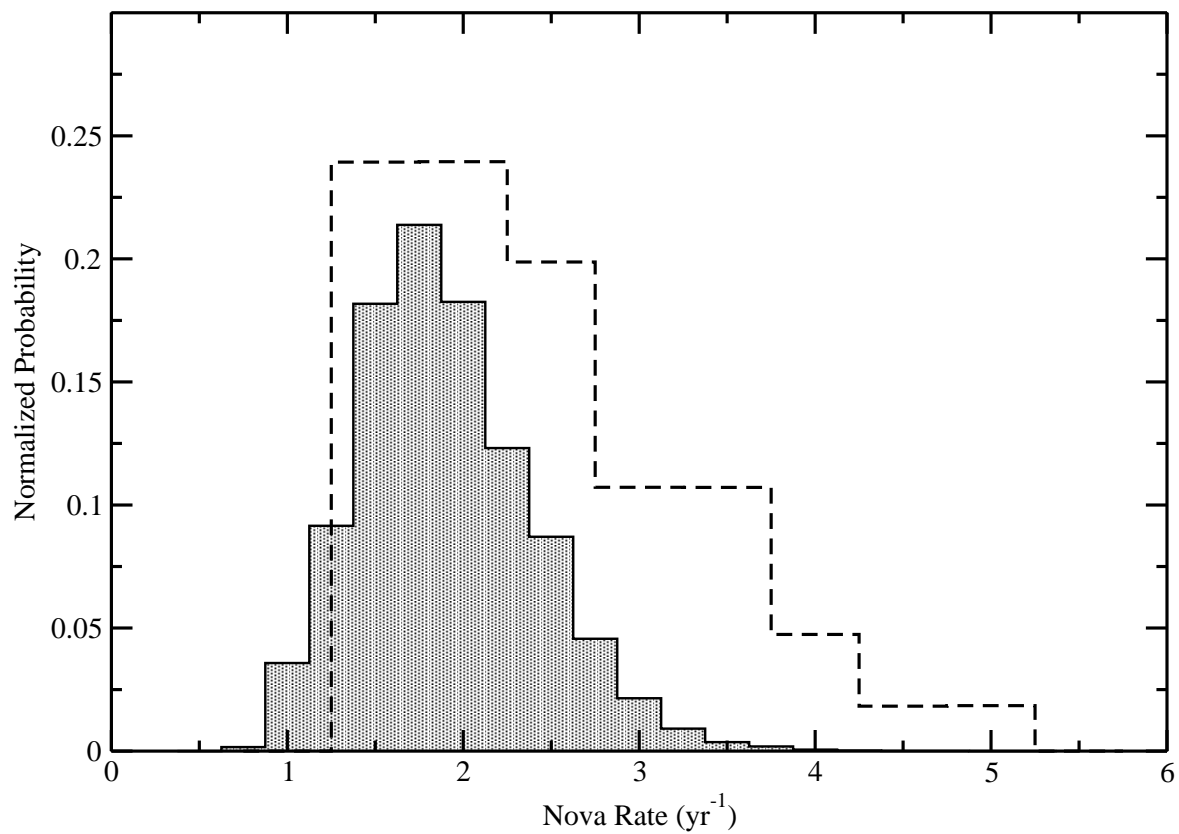}
\caption{Comparison of the results from the Monte Carlo simulation
 for the two data sets. The shaded region represents the results
 for the seven novae observed with the INT,
 with a corresponding nova rate of
 $R=1.8^{+0.6}_{-0.2}$~yr$^{-1}$. The dashed histogram represents
 the two novae discovered with the BT,
 with a peak in the probability at $R=2.0^{+0.8}_{-0.4}$~yr$^{-1}$.}
\label{fig:MC_comp}
\end{figure}

As a consistency check, we have also analyzed the BT and INT
data sets independently. Despite the small number of novae detected
in the BT survey, as shown in Figure~\ref{fig:MC_comp},
the two runs yielded consistent results with
the BT and INT surveys giving nova rates of $2.0^{+0.8}_{-0.4}$
and $1.8^{+0.6}_{-0.2}$ novae per year, respectively.

\subsection{The Mean Nova Lifetime Procedure}

In addition to our Monte Carlo calculation, we have also
made an estimate of the nova rate using a mean nova lifetime method,
which is similar to the method first employed by \cite{zwi42} to determine
extragalactic supernova rates.
This method, which has been described in several previous
efforts to determine extragalactic nova rates
\citep[e.g.,][and references therein]{sha08},
suffers from the major drawback that it requires the
adoption of a specific limiting magnitude for the survey. The method
implicitly assumes that all novae brighter than the limiting magnitude
are discovered, while all novae fainter than the limiting magnitude
are missed. This is clearly not the case as demonstrated by our
artificial star tests (see Figure \ref{fig:MonteCarlo}). In addition,
the method implicitly assumes that there is no variation
in the limiting magnitude from epoch-to-epoch.

In order to correct these limitations, we have modified the
traditional mean nova lifetime
calculation as follows. In the case of multi-epoch surveys
where the limiting magnitude varies with epoch,
the nova rate, $R$, in the surveyed region can be expressed as: 
 
\begin{equation}
R = 2.0 \times \frac{\sum_i n_i (M < M_{c,i})}{T_{e}}
\label{eqn:NovaRate}
\end{equation}

\noindent
where $n_i (M < M_{c,i})$ is the total number of novae
observed in the $i$th epoch that are brighter than the
absolute magnitude $M_{c,i}$ where 50\% of the
novae are expected to be recovered,
and $T_{e}$ is
the ``effective survey time."
The effective survey time depends both on the mean nova
lifetime, $\tau_{c,i}$, which is defined as the length of time a typical nova
remains brighter than the limiting absolute magnitude of the $i$th epoch,
and the frequency of sampling in the survey.
We have:

\begin{equation}
T_{e} = \tau_{c,1} + \sum_{i=2}^{n}min(t_{i} - t_{i - 1},\tau_{c,i})
\label{eqn:EffTime}
\end{equation}

\noindent
where $t_{i}$ is the time of the $i$th observation. Based on observations from
the bulge of M31, \citet{sha00} have provided a simple calibrated relationship
between $\tau_{c,i}$ and $M_{c,i}$, which we implement here:

\begin{equation}
\log\tau_{c,i}({\rm day}) \simeq (6.1 \pm 0.4) + (0.56 \pm 0.05)M_{c,i}
\label{eqn:TauTime}
\end{equation}

Values of $M_{c,i}$ corresponding to 50\% completeness at a given epoch
have been determined using the completeness functions
described in Section 3.1, taking
$\mu_{0} = 27.48 \pm 0.10$ \citep{fre01}, and assuming
a foreground Galactic extinction of $A_B=\sim$0.17 mag \citep{sch98}.
The effective survey time was then calculated using the survey dates
provided in Table~1, coupled with Equations (\ref{eqn:EffTime}) and
(\ref{eqn:TauTime}). We find $T_{e} = 3466 \pm 36$ days.
Noting that eight of the nine novae discovered satisfy the
condition that $M < M_{c,i}$ at the time of discovery,
Equation (\ref{eqn:NovaRate}) then yields
a nova rate of 1.7 $\pm$ 0.6 yr$^{-1}$.
Given the uncertainties inherent in this method,
the nova rate agrees reasonably well
with the value determined in the more sophisticated
Monte Carlo calculation.

\subsection{Luminosity-specific Nova Rate}

To compare nova rates between different galaxies, it is necessary
to normalize the rates by the luminosity of the galaxy. Typically,
the rates are normalized by the integrated infrared $K$-band luminosity,
which provides a better tracer of the mass in stars than does
visual light. The resulting $K$-band LSNR,
$\nu_{K}$, is usually parameterized as the number of novae
per year per 10$^{10}$ $L_{\odot,K}$.

The $K$-band luminosity of NGC 2403 has been determined
using the integrated $K$-band magnitude of $6.19\pm0.04$
measured in the Two Micron All Sky Survey's (2MASS)
Large Galaxy Atlas \citep{jar03}. At the distance of N2403,
a nova rate of 2.0$^{+0.5}_{-0.3}$ yr$^{-1}$ yields
$\nu_K = (2.5\pm0.7) \times 10^{-10}L_{\odot,K}$ yr$^{-1}$.
The LSNR of NGC 2403 is typical of those of other galaxies with
measured nova rates~\citep[e.g., see][]{gue10}, and is
consistent with that of the similar Hubble type galaxy, M33,
where $\nu_K = (2.17\pm0.89) \times 10^{-10}L_{\odot,K}$ yr$^{-1}$
\citep{wil04}.

\subsection{The Nova Spatial Distribution}

\begin{figure}
\includegraphics[angle=0,scale=0.80]{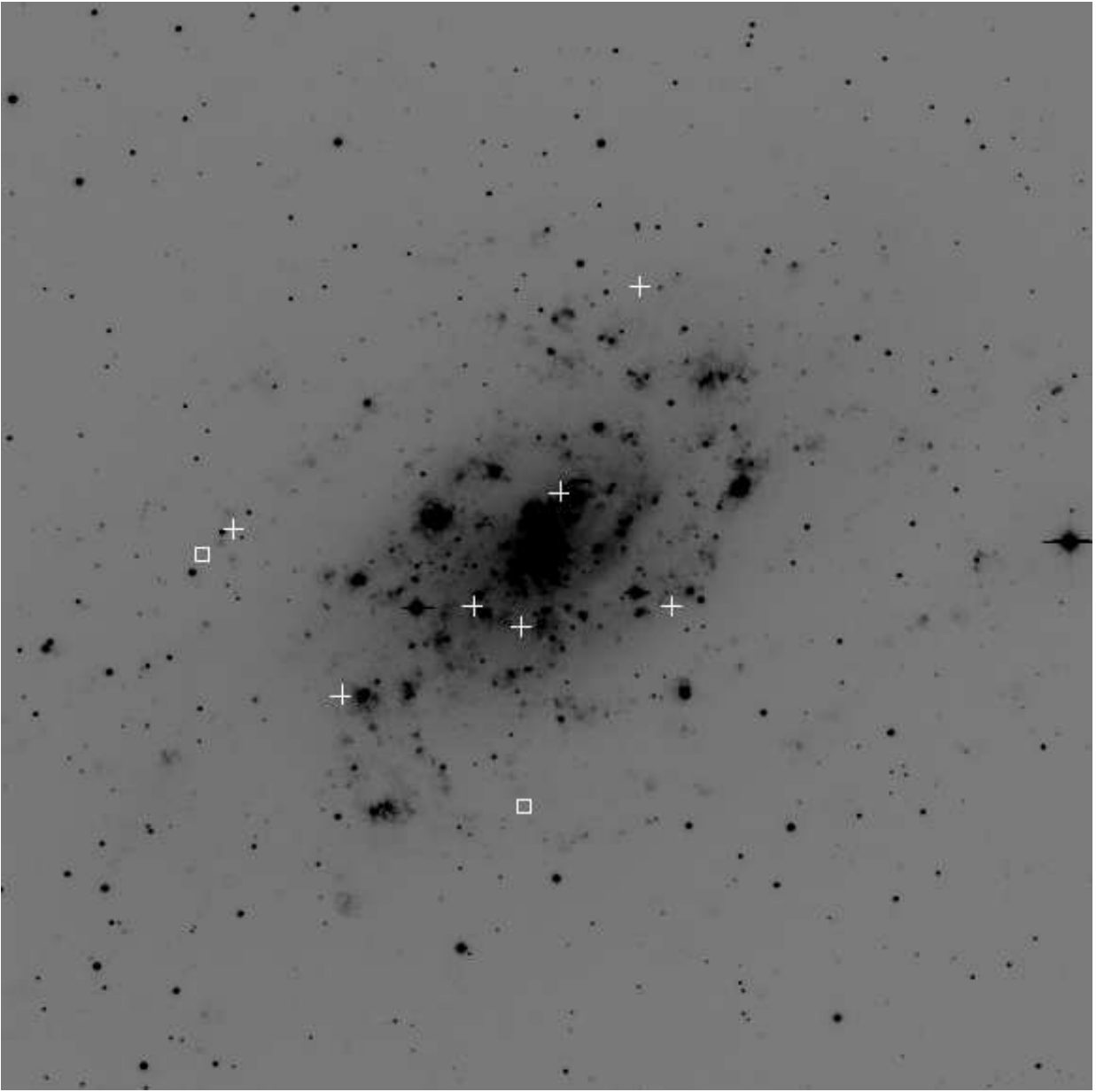}
\caption{Spatial distribution of the nine nova candidates
  found in NGC 2403 plotted over the 2007 March 11
  image of the galaxy from the BT survey. The squares
  indicate the positions of the two novae discovered in the BT survey.
  The image is approximately 18$'$ on a side with N up and E left.}
\label{fig:SpatialDistr}
\end{figure}

\begin{figure}
\includegraphics[angle=-90,scale=0.65]{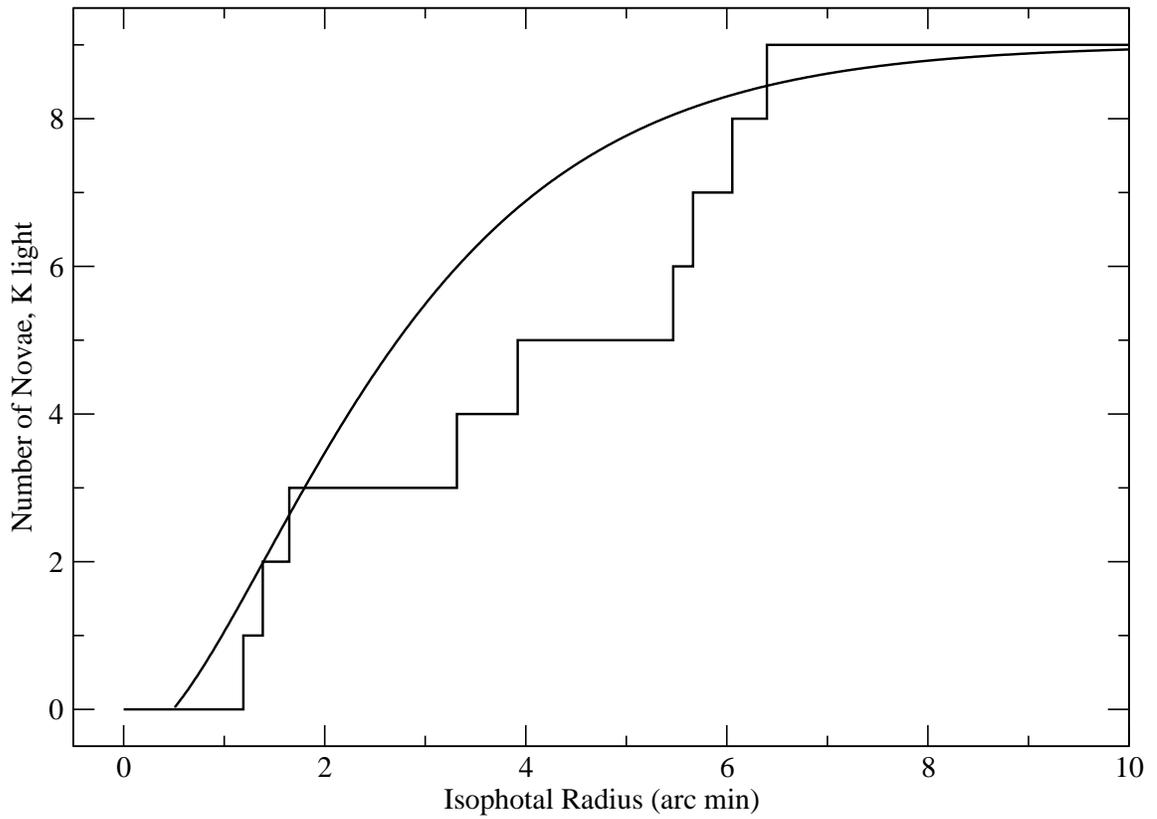}
\caption{Cumulative nova distribution compared with the
 integrated $K$-band light (smooth curve)
 of \citet{pie97}. A K-S test (KS = 0.25)
 suggests that the background galactic
 light falls off faster than the observed nova distribution.}
\label{fig:NovaLightDistr}
\end{figure}

In Figure \ref{fig:SpatialDistr}, we have
plotted the positions of all nine nova candidates discovered in
our overall survey over an image of the galaxy from the BT observations.
To further explore the spatial distribution of the novae,
we have used the $K$-band photometry of \citet{pie97} to
compute the radii of the isophotes that pass through the
nova positions.
The cumulative distribution of the nova isophotal radii is compared with
the cumulative distribution of the galaxy's $K$-band light in
Figure \ref{fig:NovaLightDistr}.
The integrated $K$-band light falls off somewhat more rapidly than the
observed nova distribution.
A Kolmogorov-Smirnov (K-S) test shows that there is a 25\% chance that
the distribution would differ by as much as observed if they were drawn
from the same parent distribution (i.e., if the nova frequency was
proportional to the surface brightness of the galaxy). Thus, it is possible that
the nova distribution is incomplete in the inner regions of the galaxy
where the background is particularly bright and inhomogeneous.

\section{Discussion}

The nova survey of NGC 2403 is a continuation of an effort to
understand the role that the host galaxy morphology has on its nova rate.
Previous surveys have yielded conflicting results with
\citet{del94} arguing that the LSNR is a function of
Hubble type (with late-type galaxies being more prolific nova
producers), while \citet{cia90a} and many subsequent studies
\citep[e.g.,][]{sha00,wil04,coe08,gue10} have found that the
LSNR is insensitive to Hubble type. Exceptions seem to be
the Large Magellanic Cloud (LMC) and the Small Magellanic Cloud (SMC),
where the LSNRs appear to be elevated by a factor of 2--3
when compared with other galaxies for which nova rates
have been measured. Their relatively low luminosities (and correspondingly
low absolute nova rates),
renders the LSNRs for the LMC and SMC particularly uncertain.
More work needs to be done on these galaxies
before we can be confident in their LSNRs.

Among the late type spiral galaxies, M33 \citep[SA(s)cd;][]{dev91}
has been particularly noteworthy. \citet{wil04} found a nova rate of
$2.5^{+1.0}_{-0.7}$ per year while \citet{del94} found a rate,
$4.6\pm0.9$, that was almost a factor of two higher. This difference
resulted in \citet{del94} finding an elevated LSNR for M33 that was comparable
to that of the LMC, while the value determined by \citet{wil04},
$2.17\pm0.89$ novae per year per $10^{10}L_{\odot,K}$,
was typical of most other galaxies which tend to lie in a range of
1--3 novae per year per $10^{10}L_{\odot,K}$.
NGC 2403 \citep[SAB(s)cd;][]{dev91} is interesting in that it represents
a near morphological clone of M33. Our estimate of
$\nu_K = (2.5\pm0.7) \times 10^{-10}L_{\odot,K}$ yr$^{-1}$ for NGC 2403
is typical of other galaxies, and is consistent with the value
for M33 determined by \citet{wil04}.

It is worth noting that
the spatial distribution of the novae in NGC 2403 does not
uniformly follow the background $K$-band light distribution of the galaxy,
as shown by Figures~\ref{fig:SpatialDistr} and \ref{fig:NovaLightDistr}.
The light distribution falls off faster than the nova distribution,
which suggests either that novae are more frequent in the outer disk
regions, or, that we may be missing more novae in the brighter
central region of the galaxy. The latter possibility is more likely
given that our artificial star tests show that the completeness is
reduced somewhat at small isophotal radii. Since the completeness
over the entire galaxy is taken into account by our Monte Carlo
simulation, our derived LSNR is not affected.

A potential source of error implicit in our analysis concerns
our assumption that the H$\alpha$ light curves used in the
Monte Carlo simulation (which are based on novae observed in M31 and M81)
are characteristic of novae in NGC 2403. It is possible that
the nova speed classes are a function of stellar population.
If, for example, the
novae in NGC 2403 are ``faster" than those observed in M31 and M81,
as has been suggested for the LMC \citep[e.g.,][]{del93,cap90},
or if there is a significant population of ``faint and fast" novae
in M31 \citep[e.g.,][]{kas11} that are not
represented in our model light curves, then our Monte Carlo simulations
would underestimate the true nova rate in NGC 2403
(i.e., a lower assumed intrinsic nova rate in a population
of ``slower" novae would suffice to produce the observed number of novae).
That said, given the fact that two of the nine novae observed
in NGC~2403 faded particularly slowly, and were visible for over 6 months,
it is clear that the NGC 2403 novae are unlikely to be significantly
faster, on average, compared with those seen in M31 and M81.

\section{Conclusions} 

Over the course of 48 nights spanning more than a decade,
we have identified a total of nine nova candidates in NGC 2403.
Two of the candidates faded from peak brightness slowly,
but were not seen in prior epochs and did not reappear in any later epochs.
Based on a Monte Carlo simulation,
our best estimate for the nova rate of NGC 2403
is 2.0$^{+0.5}_{-0.3}$ yr$^{-1}$.
The observations were taken using two telescopes, the 2.3-m BT and
the 2.5-m INT,
which yielded a total of 12 and 36 nights of observation, respectively.
If analyzed separately, our estimate of the nova rate
for the INT is $R=1.8^{+0.6}_{-0.2}$~yr$^{-1}$ and
$R=2.0^{+0.8}_{-0.4}$~yr$^{-1}$ for the BT,
each of which are consistent with our combined nova rate.
Based on the integrated $K$ magnitude from the 2MASS survey,
we estimate the LSNR to be
$\nu_K = 2.5\pm0.7$ novae per year per $10^{10}L_{\odot,K}$,
which is consistent with that of the similar Hubble type galaxy, M33.

Given the sensitivity of derived nova rates to the photometric
evolution of novae (e.g., peak magnitude and rate of decline),
future studies of extragalactic nova rates should focus on
measuring light curve parameters for novae in galaxies
spanning a range of Hubble types. Such studies will require
frequent sampling over an extended period of time, and will be
facilitated tremendously by observations made with current
and planned survey telescopes such as PanStarrs
and the LSST.

\acknowledgements

This research has made use of the NASA/IPAC Extragalactic Database
(NED) which is operated by the Jet Propulsion Laboratory,
California Institute of Technology,
under contract with the National Aeronautics and Space Administration
and the SIMBAD database,
operated at CDS, Strasbourg, France. 
We thank Raine Karjalainen for making it possible to use the Isaac Newton
Telescope images taken during three D-nights in 2011-2012.
This paper makes use of data obtained from the Isaac Newton Group Archive
which is maintained as part of the CASU Astronomical Data Centre
at the Institute of Astronomy, Cambridge.
Based in part on data collected at Subaru Telescope and obtained
from the SMOKA, which is operated by the Astronomy Data Center,
National Astronomical Observatory of Japan.
Work of K.H. was supported by the project RVO:67985815.
A.W.S. thanks the NSF for support through grant AST 1009566.

\clearpage
\begin{deluxetable}{c c c c c}
\tabletypesize{\scriptsize}
\tablewidth{0pt}
\tablecolumns{5}
\tablecaption{Summary of Observations\tablenotemark{a}}
\tablehead{
\colhead{} & \colhead{Julian Date} & \colhead{Number of} & \colhead{Total
  Integration Time} & \colhead{} \\ \colhead{UT Date} & \colhead{(2,450,000+)} &
\colhead{Exposures} & \colhead{(hr)} & \colhead{$m_{lim}$}}
\startdata
2001 Feb 21 & 1962.417  & 1 & 0.67 &  22.3 \\
2002 Dec 30 & 2638.624  & 1 & 0.33 &   23.1 \\
2004 Jan 30 & 3035.464  & 1 & 0.5 &  22.3 \\
2004 Feb 12 & 3048.420  & 1 & 0.5 &  21.9 \\
2005 Jan 01  & 3385.449  & 1 & 0.08 &  21.5 \\
2005 Mar 20 & 3450.392 & 1 & 0.25 &  22.2 \\
2005 Oct 30\tablenotemark{b} & 3673.860 & 10 & 2.5 &  22.1 \\
2005 Oct 31\tablenotemark{b} & 3675.006 & 2 & 0.5 &  22.1 \\
2005 Dec 28\tablenotemark{b} & 3732.840 & 14 & 3.5 & 23.1 \\
2006 Feb 6\tablenotemark{b} & 3772.679 & 8 & 2.0 & 23.3 \\
2006 Feb 7\tablenotemark{b} & 3773.690 & 2 & 0.42 & 23.3 \\
2006 Feb 21\tablenotemark{b} & 3787.801 & 6 & 1.5 & 22.7 \\
2006 Feb 23\tablenotemark{b}  & 3789.641 & 4 & 1.0 & 22.7 \\
2006 Mar 11 & 3806.409  & 1 & 0.5 & 22.7 \\
2006 Mar 14 & 3809.447  & 1 & 0.55 & 22.0 \\
2006 Apr 19\tablenotemark{b} & 3844.668 & 8 & 2.0 & 22.4\\
2006 Oct 30 & 4038.743  & 1 & 0.13 & 20.5  \\
2006 Nov 28\tablenotemark{b} & 4067.874 & 10 & 2.0 & 22.9 \\
2006 Dec 13\tablenotemark{b} & 4082.800 & 15 & 2.5 & 23.1 \\
2006 Dec 27& 4096.619  & 1 & 0.56 & 22.1\\
2007 Jan 01 & 4103.673  & 1 & 0.33 & 20.8 \\
2007 Jan 04 & 4104.554  & 1 & 0.33 & 20.7 \\
2007 Jan 06 & 4106.638  & 1 & 0.33 & 20.9 \\
2007 Jan 09 & 4109.618  & 1 & 0.33 & 19.7 \\
2007 Jan 27\tablenotemark{b} & 4127.683 & 14 & 2.33 & 23.2 \\
2007 Mar 11\tablenotemark{b} & 4170.652 & 18 & 3.0 & 22.7 \\
2007 Sep 29 & 4372.733  & 1 & 0.33 & 21.9 \\
2007 Sep 30 & 4373.726  & 1 & 0.33 & 21.2 \\
2008 Jan 28 & 4494.490  & 2 & 1.0 & 23.7 \\
2008 Jan 30 & 4495.579  & 3 & 1.25 & 20.8 \\
2008 Oct 12 & 4751.685  & 2 & 1.0 & 22.7 \\
2008 Oct 13 & 4752.667  & 1 & 0.5 & 21.4 \\
2008 Oct 25 & 4764.700 & 1 & 0.34 & 21.6 \\
2008 Oct 26 & 4765.744  & 1 & 0.17 & 21.2 \\
2008 Oct 28 & 4767.729  & 2 & 0.79 & 23.0 \\
2008 Nov 18 & 4788.624  & 2 & 1.0 & 22.4 \\
2009 Jan 19 & 4851.453  & 2 & 1.0 & 22.3 \\
2009 Jan 21 & 4852.609  & 2 & 1.0 & 22.8 \\
2009 Jan 22 & 4853.610  & 1 & 0.5 & 22.6 \\
2009 Feb 17 & 4880.443  & 1 & 0.17 & 22.0 \\
2009 Feb 19 & 4882.392  & 1 & 0.17 & 21.3 \\
2009 Apr 11 & 4933.402  & 2 & 1.0 & 23.0 \\
2009 Apr 12 & 4934.404  & 2 & 1.0 & 23.4 \\
2010 Feb 5  & 5233.443  & 2 & 1.0 & 23.6 \\
2010 Sep 20 & 5459.709 & 2 & 1.0 & 22.8 \\
2011 Mar 23 & 5644.456  & 1 & 0.08 & 21.6 \\
2012 Jan 21 & 5947.709  & 1 & 0.5 & 22.4 \\
2012 Apr 28 & 6046.416  & 1 & 0.5 & 22.0 \\
\enddata
\tablenotetext{a}{Unless otherwise noted,
 observations were taken with the 2.5-m INT.} 
\tablenotetext{b}{Data obtained with the 2.3-m BT.}
\end{deluxetable}

\clearpage
\begin{deluxetable}{cccccc}
\tabletypesize{\scriptsize}
\tablewidth{0pt}
\tablecolumns{6}
\tablecaption{NGC 2403 Nova Candidates}
\tablehead{
\colhead{} & \colhead{Julian Date} & \colhead{$\alpha$} & \colhead{$\delta$} &
\colhead{$a$} & \colhead{$m_{H \alpha}$}  \\ \colhead{Nova} & \colhead{(2,450,000+)} & \colhead{(J2000.0)} & \colhead{(J2000.0)} &
\colhead{(arcmin)} & \colhead{(mag)}}
\startdata
N2403 2001-02a & 1962.417 & 07 37 21.90 & 65 33 49.7 & 3.92 & 20.9 \\
              & 2638.624 & \nodata & \nodata & \nodata & 22.1 \\
N2403 2002-12a\tablenotemark{a} & 2638.624 & 07 37 38.36 & 65 36 29.8 & 6.05 & 18.3 \\
			  & 2638.635   & \nodata & \nodata & \nodata & 20.9\tablenotemark{b} \\
			  & 2643.5    & \nodata & \nodata & \nodata & 21.4\tablenotemark{c} \\			                
N2403 2005-01a\tablenotemark{a} & 3385.449 & 07 36 53.98 & 65 34 54.7 & 1.65 & 19.3 \\
			  & 3450.392  & \nodata & \nodata & \nodata & 19.7 \\
			  & 3806.409 & \nodata & \nodata & \nodata & 21.9 \\
			  & 3808.453  & \nodata & \nodata & \nodata & 21.8 \\
N2403 2005-10a\tablenotemark{d} & 3673.860 & 07 37 43.07 & 65 36 05.4 & 6.40 & 21.8 \\
			  & 3732.840  & \nodata & \nodata & \nodata & 23.1 \\		  
N2403 2005-12a\tablenotemark{d} & 3732.840 & 07 36 53.79 & 65 32 03.5 & 5.66 & 21.0 \\
			  & 3772.679  & \nodata & \nodata & \nodata & 21.4 \\
			  & 3787.801 & \nodata & \nodata & \nodata & 21.6 \\			  
N2403 2007-09a & 4372.733  & 07 36 47.72 & 65 37 04.0 & 1.19 & 19.6 \\
			  & 4373.726  & \nodata & \nodata & \nodata & 19.7 \\
N2403 2008-01a & 4494.490  & 07 37 01.33 & 65 35 14.7 & 1.38 & 20.7 \\
			  & 4495.579  & \nodata & \nodata & \nodata & 20.9 \\
N2403 2009-03a\tablenotemark{e} & 4933.402  & 07 36 35.04 & 65 40 21.4 & 5.47 & 19.2 \\
			  & 4934.404 & \nodata & \nodata & \nodata & 19.1 \\
N2403 2010-02a & 5233.443  & 07 36 30.57 & 65 35 14.3 & 3.31 & 19.4 \\
			  & 5459.709  & \nodata & \nodata & \nodata & 22.0 \\
\enddata
\tablecomments{The units for right ascension are hours, minutes, and seconds.
  The units for declination are degrees, arcminutes, and arcseconds.  We
  assume these values to be accurate to $\sim 1\arcsec$. The parameter
  $a$ is the isophotal radius of the nova.
  Unless otherwise noted, nova were identified using the 2.5-m INT.} 
\tablenotetext{a}{Discovery first reported in \cite{hor08}.}
\tablenotetext{b}{r-band magnitude taken with the 2.5-m INT with the WFC instrument}
\tablenotetext{c}{R-band magnitude taken with the 0.41-m telescope at Kitt Peak with the ST-10XME instrument}
\tablenotetext{d}{Nova found using the 2.3-m BT with the P90 camera.} 
\tablenotetext{e}{Nova discovered independently by P60-FasTING \citep{kas09}}
\end{deluxetable}

\clearpage

\clearpage

\clearpage

\clearpage

\clearpage
\end{document}